\documentclass[lnbip]{svmultln}
\usepackage{graphicx}
\usepackage{booktabs}
\usepackage{paralist}
\usepackage{todonotes}
\usepackage{makeidx}  
\begin{document}
\mainmatter              
\title{Empirical Research Plan: Effects of Sketching  \\ on Program Comprehension}
\titlerunning{Effects of Sketching on Program Comprehension}  
%
\author{Sebastian Baltes\inst{1} \and Stefan Wagner\inst{2}}
\authorrunning{Sebastian Baltes, Stefan Wagner}   
%
%
\institute{
University of Trier, Germany\\
\email{research@sbaltes.com}
\and
University of Stuttgart, Germany\\
ORCID ID: 0000-0002-5256-8429\\ 
\email{stefan.wagner@informatik.uni-stuttgart.de}}

\maketitle              

\begin{abstract}        
Sketching is an important means of communication in software
engineering practice.
Yet, there is little research investigating the use of sketches.
We want to contribute a better understanding of sketching, in particular
its use during program comprehension.
We propose a controlled experiment to investigate the effectiveness and
efficiency of program comprehension with the support of sketches
as well as what sketches are used in what way.
\keywords {experiment, sketching, program comprehension}
\end{abstract}
\section{Introduction}
Software is inherently abstract and has no natural representation except source code.
Thus, especially for program comprehension, visualizations are important~\cite{Storey06}. 
Sketches are an example for informal visualizations that are often created when understanding or explaining source code~\cite{Baltes14}.  
In the past, however, these informal artefacts did not get the amount of attention by the software engineering research community that their relevance in software development practice could imply.
With our proposed study, we want to analyse if and how sketching improves program comprehension when explaining source code.
Furthermore, we want to gain a better understanding of what sketches are used in what way to explain the source code.
In the description of our experiment, we follow the guidelines of Jedlitschka, Ciolkowski and Pfahl~\cite{Jedlitschka05}.

\section{Related Work}

One of the main purposes of sketching in software development is communication~\cite{Baltes14, Cherubini07}.
To this end, developers often employ ad hoc notations that rarely adhere to standards like the \emph{Unified Modeling Language} (UML)~\cite{Baltes14, Petre13}.
The ambiguity in sketches is a source of creativity~\cite{Goldschmidt03} and they support problem solving and understanding~\cite{Suwa02}.
In other areas like engineering, controlled experiments have shown that the possibility to sketch has a positive effect on the quality of the solutions~\cite{Schuetze03}.
In our study, we want to analyse if sketches improve program comprehension in a setting where one developer explains a piece of source code to a colleague.
To be able to compare the effect of sketching on program comprehension, we measure task correctness and response time~\cite{Dunsmore00}.

\section{Experiment Planing}


The overall goal of our research is to better understand the use and usefulness of sketches in software engineering.
In this experiment, we especially focus on sketching as a means of program comprehension in the communication
between two developers. The goal of our experiment is:\\
\textbf{Analyze} sketching while explaining source code\\
\textbf{for the purpose of} evaluating its impact on program comprehension\\
\textbf{with respect to} its effectiveness and efficiency\\
\textbf{from the viewpoint of} the developer\\
\textbf{in the context of} the conference XP 2016.\\

From this, we derive three research questions. The first two are more descriptive and
exploratory to better understand which sketches developers use and how they use them
while explaining source code to another developer. The third covers then the causal
relationship of using sketches onto the effectiveness and efficiency of comprehending
the source code.

\noindent\textbf{RQ~1}: Which sketches do developers use to explain code? \\ 
\textbf{RQ~2}: How do developers explain code with and without sketches?\\ 
\textbf{RQ~3}: How does the effectiveness and efficiency of the understanding of the code differ 
when it was explained with or without a sketch? \\


\subsection{Experimental Units and Materials}

The participants of the experiments will be pairs of developers. They will explain source
code to each other. They have to be professional software developers.


We will use four different small open-source software systems in commonly known programming
languages such as Java or C\#. As the developers do not know the source code beforehand but have
to explain them, we limit the systems to 500 LOC at most.

\subsection{Tasks}

The basic task for each pair of developers is to understand the source code of a small software
system and then explain certain aspects to each other. The source code will be made available
on an iPad. In case they should sketch, this will be done on paper. The aspects to explain will
be low-level and code-centric. Afterwards, the developer the aspect was explained to, will answer
questions evaluating how well they understood the explanations.


\subsection{Hypotheses, Parameters and Variables}

The central independent variable of the experiment is the use of sketching. The dependent
variables we are going to measure are the time needed until the explained aspect is understood
and the correctness of the understanding. For the explorative part, we also document which
types of sketches (e.g.\ different UML diagrams) they used and how they themselves judged
the difference in explanations.

The two null hypotheses we are going to investigate are:\\
$\mbox{H}_{01}$: \emph{There is no difference in the effectiveness of comprehension with or without sketches.}\\
$\mbox{H}_{02}$: \emph{There is no difference in the efficiency of comprehension with or without sketches.}

Furthermore, we will document further context variables such as the experience of the developers
with the programming languages and whether they have previously worked together.

\subsection{Experiment Design}

We will employ a blocked and balanced design. Hence, from each developer pair, the first developer will first read and 
explain a software system with sketching and then read another software system and explain it without sketching. The second
developer will do the same but first without sketching and then with sketching.

We will openly invite the XP 2016 participants to join the experiment in pairs. Therefore, the sample is a convenience sample.

\subsection{Procedure}

We need a separate location for the experiment so that the participants can concentrate on understanding
and explaining. We could hold it as one event during the conference or continuously over the whole conference
depending on the fit to the conference schedule. We will put up lists in which the developers can volunteer to
participate.

The first step when a pair starts the experiment is that they receive an iPad each with their two software systems
to explain together with the question concerning the aspect they later have to explain to the other developer. Then (step~2) both get
time to read the first system. In step 3, participant 1 explains the first system to participant 2 without a sketch. The time
for this is measured on the iPad. Step 4 is a short questionnaire for participant 2 to check the correctness of their 
understanding. In step 5, participant 2 explains their software system aspect to participant 1 with the help of sketches
on provided paper (including time measurement on the iPad). In step 6, participant 1 answers the short questionnaire 
concerning correctness.

Next, in step 7, both participants read the next question and source code. Then, the same procedure is repeated but
participant 1 gets to use sketches while participant 2 does not. We will ask about the general experience and context
factors in a final questionnaire. 

\subsection{Analysis Procedure}

We will analyse the quantitive data to test the two hypotheses using an ANOVA analysis (RQ 3). Furthermore, we will
qualitatively analyse the sketches and the answers to the open questions in the final questionnaire (RQ 1 and 2).

\section{Summary and Future Work}

In summary, we want to conduct a controlled experiment to better understand how developers use sketches
in explaining source code as well as the effects on effectiveness and efficiency of the comprehension. The results
of the experiment allow us to reduce the discrepancy between research concentrating on more formally defined
modelling languages and the relevance of sketching in practice. Furthermore, we want to use the gained insights
to work on a sketching language and tool support to aid practitioners in sketching in an efficient and effective way.

%
\bibliographystyle{splncs}
\bibliography{literature} 

\newpage
\footnotesize{
\noindent This chapter is distributed under the terms of the Creative Commons Attribution-NonCommercial 4.0 International License 
(\url{http://creativecommons.org/licenses/by-nc/4.0/}), which permits any noncommercial use, duplication, adaptation, distribution and reproduction in any medium or format, as long as you give appropriate credit to the original author(s) and the source, a link is provided to the Creative Commons license and any changes made are indicated.}

\end{document}